\documentclass[a4paper,11pt]{article}
\pdfoutput=1 

\usepackage{jinstpub} 
\usepackage[abs]{overpic}
\usepackage{subfigure}
\usepackage{lineno}

\title{\boldmath Simulations of CMOS pixel sensors with a small collection
electrode, improved for a faster charge collection and
increased radiation tolerance}

\author[a]{M. Munker,\note{Corresponding author.}}

\author[b]{M. Benoit,}
\author[a]{D. Dannheim,}
\author[c]{A. Fenigstein,}
\author[a]{T. Kugathasan,}
\author[c]{T. Leitner,}
\author[a]{H. Pernegger,}
\author[a]{P. Riedler,}
\author[a]{W. Snoeys}

\affiliation[a]{CERN,\\Geneva 23, Switzerland}
\affiliation[b]{University of Geneva,\\Geneva 4, Switzerland}
\affiliation[c]{TowerJazz Semiconductor,\\Migdal Haemek, 23105, Israel}

\emailAdd{magdalena.munker@cern.ch}

\abstract{
CMOS pixel sensors with a small collection electrode combine the advantages of a small sensor capacitance with the advantages of a fully monolithic design.
The small sensor capacitance results in a large ratio of signal-to-noise and a low analogue power consumption, while the monolithic design reduces the material budget, cost and production effort.
However, the low electric field in the pixel corners of such sensors results in an increased charge collection time, that makes a fully efficient operation after irradiation and a timing resolution in the order of nanoseconds challenging for pixel sizes larger than approximately forty micrometers.
This paper presents the development of concepts of CMOS sensors with a small collection electrode to overcome these limitations, using three-dimensional Technology Computer Aided Design simulations.
The studied design uses a $\mathrm{0.18 \, \mu m}$ process implemented on a high-resistivity epitaxial layer.

}


\arxivnumber{1903.10190} 

\proceeding{9$^{\text{th}}$ Workshop on Semiconductor Pixel Detectors for Particles and Imaging (PIXEL)\\
  December 2018\\
  Taipei}

\begin{document}
\maketitle
\flushbottom
\section{Introduction}

Monolithic pixel-detector technologies reduce the production effort and cost while reducing the material budget in tracking systems of detectors of high-energy physics experiments.
Integrated CMOS pixel sensors with a small collection electrode offer a small sensor capacitance, a favourable signal-to-noise ratio and power consumption, and the potential for excellent spatial and timing resolution~\cite{walter_power}. 
Such sensors have been developed and adopted for the ALICE ITS upgrade using a standard $\mathrm{0.18 \, \mu m}$ CMOS imaging sensor process on a high resistivity epitaxial layer~\cite{alice_its}. 
Modifying the process to achieve full depletion in the sensor~\cite{walter_paper} improves the radiation tolerance, of importance for the ATLAS ITk High-Luminosity upgrade~\cite{atlas_hl}, as well as the timing resolution, relevant for  the CLIC tracking system~\cite{clic0,clic1,clic2}. 
However, the electric field in the sensor reaches a minimum in the pixel corners resulting in a degraded timing resolution and efficiency loss after irradiation~\cite{bojan,ivan,roberto}. 
This is more pronounced for larger pixel sizes, and achieving full efficiency and a few ns timing resolution has been proven to be challenging for pixel sizes around $\mathrm{40 \, \mu m}$ or larger. 
This paper presents a study of two improvements of the pixel design in this modified process, a mask change and an additional implant, to further reduce the charge collection time, and therefore improve radiation tolerance and timing resolution while maintaining the small collection electrode and its benefits. The two approaches have been studied using three-dimensional self-consistent transient Technology Computer Aided Design simulations (TCAD~\cite{tcad_website}) both for non-irradiated and irradiated sensors, and have been implemented in prototype run for the ATLAS experiment~\cite{bojan,roberto}.

\section{Standard and modified process}
\begin{figure}[!ht]
\centering
	\vspace{0.4cm}
\begin{minipage}[t]{0.47\textwidth} 
\begin{overpic}[width=1\textwidth]{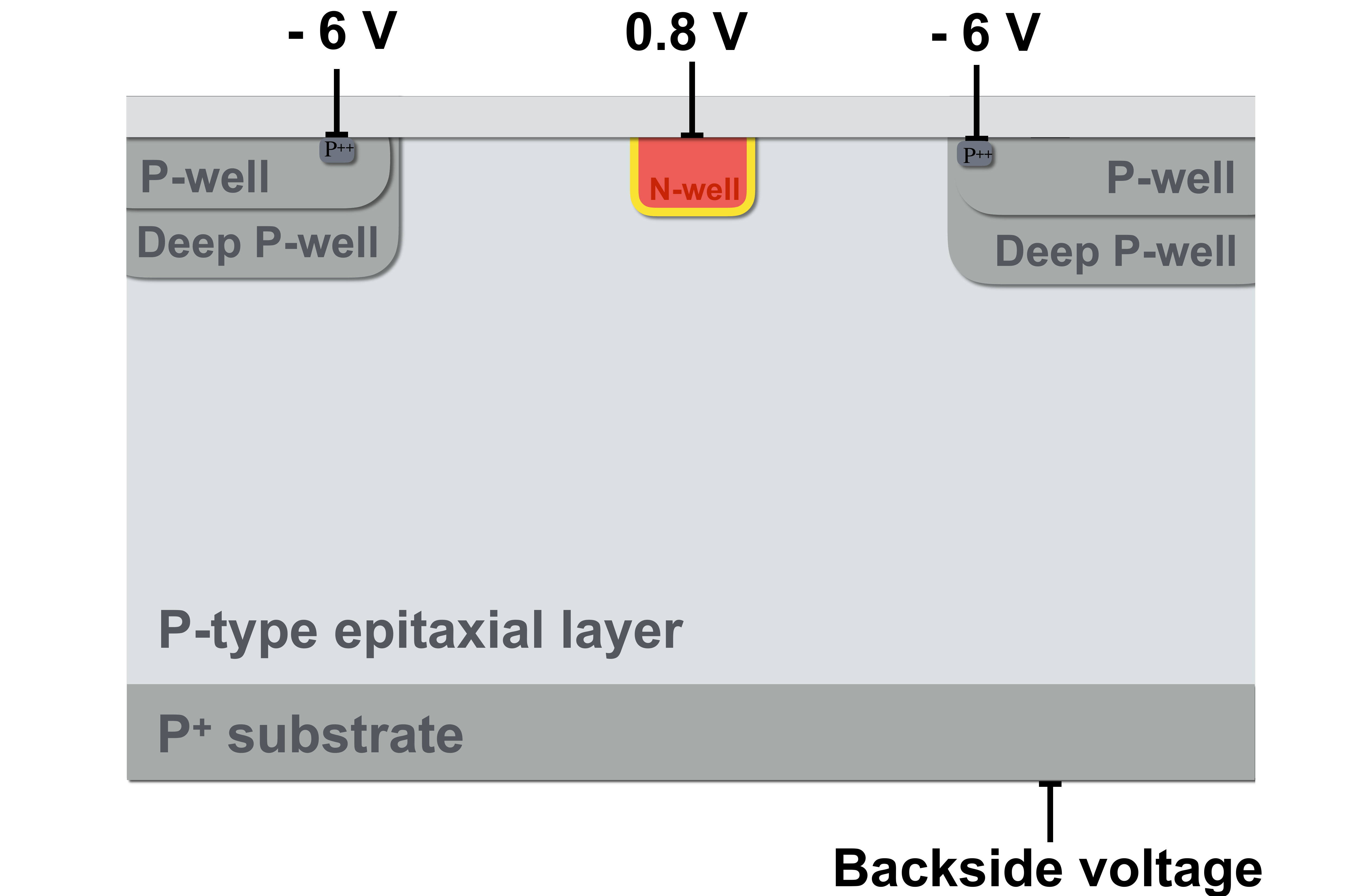}
	\put(18,139){\rotatebox{0}{ \scriptsize \textbf{Standard process:}}}
\end{overpic}
\end{minipage}
\hspace{0.3cm}
\begin{minipage}[t]{0.47\textwidth} 
\begin{overpic}[width=1\textwidth]{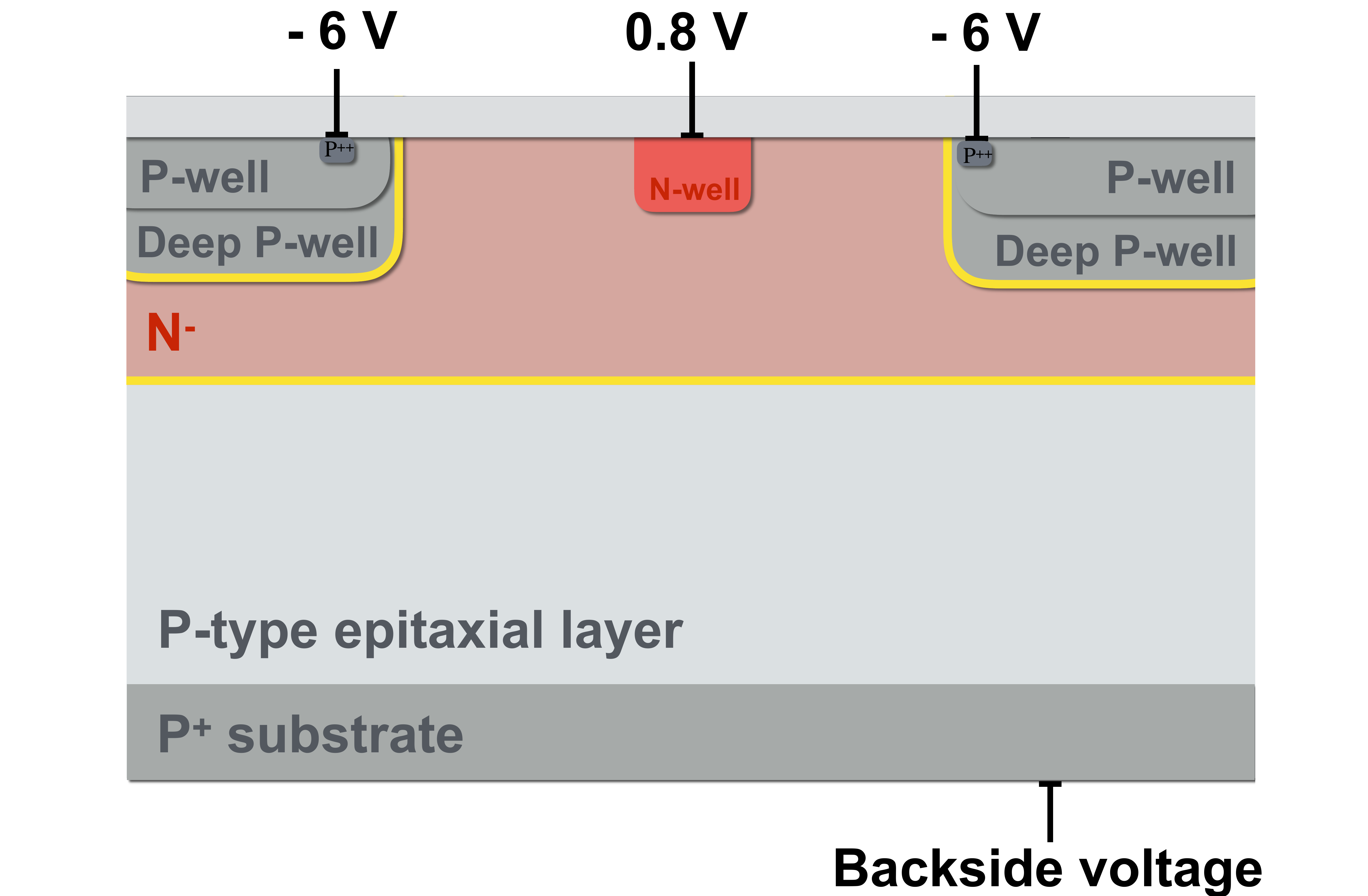}
	\put(18,139){\rotatebox{0}{ \scriptsize \textbf{Modified process:}}}
\end{overpic}
\end{minipage}
\caption{Schematic cross section (not to scale) of the CMOS standard (left) and modified (right) process with a small collection electrode. 
The implants of the CMOS circuitry are not shown. 
The yellow lines indicate the junctions.}
\label{fig:process_mod}
\end{figure}

A $\mathrm{0.18 \, \mu m }$ CMOS imaging process with a small collection electrode has been studied, as sketched in Figure~\ref{fig:process_mod}.

Full CMOS circuitry is placed inside p-wells and n-wells shielded by a deep p-well implant. 
All implants are placed on a high resistivity epitaxial layer that is grown on a low resistivity backside substrate to maximise the depleted region in the sensor.
In this standard process (see left side of Figure~\ref{fig:process_mod}), it is difficult to make the depletion layer extend from the junction around the small collection electrode laterally in the epitaxial layer between deep p-well and substrate, especially if the readout circuitry occupies a large fraction of the pixel area. With a deep low-dose n-type implant to create a planar junction under the existing implants (see right plot of Figure~\ref{fig:process_mod}), full depletion of the epitaxial layer is much easier to achieve as the depletion starts at the junction and therefore extends over the full pixel area even with low reverse bias~\cite{walter_paper,my_thesis}.

The concept of moving the junction from a small area around the collection electrode to a larger area deeper in the sensor has been pursued 
in developments to combine full depletion with a small collection electrode in monolithic sensors, both for bulk or epitaxial layer technologies~\cite{thesis_walter,goji_paper}, as well as for Silicon on Insulator (SOI) technologies~\cite{soi_paper}.

\section{Low electric field sensor regions}

In the fully depleted sensitive layer of the modified process charge collection is governed by drift, and hence by the direction and magnitude of the electric field. However, as will be shown by the three-dimensional TCAD simulations, these sensors with a small collection electrode exhibit a very non-uniform electric field, dropping to zero at the pixel corners. 

For the simulation constant voltages were applied to the different electrodes (collection electrode, p-well, substrate) in the silicon structure using ideal contacts and ideal voltage sources. 
Using this approach, the signal current produced by the sensor is absorbed by these voltage sources without charging up the capacitance associated with these electrodes. This is an ideal or best case allowing to study ultimate limitations on sensor timing performance. 
In practice, a real front end circuit does not have a zero input impedance like an ideal voltage source, and some charging of the sensor capacitance will happen. Noise contributions from sensor (shot noise) and readout circuit will also degrade timing performance as these effectively introduce random signal fluctuations.

If not mentioned otherwise, the simulations discussed in the following have been performed with a voltage of $\mathrm{0.8 \, V}$ on the collection electrode and $\mathrm{- \, 6 \, V}$ on the p-wells and backside substrate.
Cuts through the pixel centre of the simulated three-dimensional pixel cell are presented.

\begin{figure}[!ht]
\centering
\vspace{0.3cm}
\begin{minipage}[t]{1\textwidth} 
\begin{overpic}[width=1\textwidth]{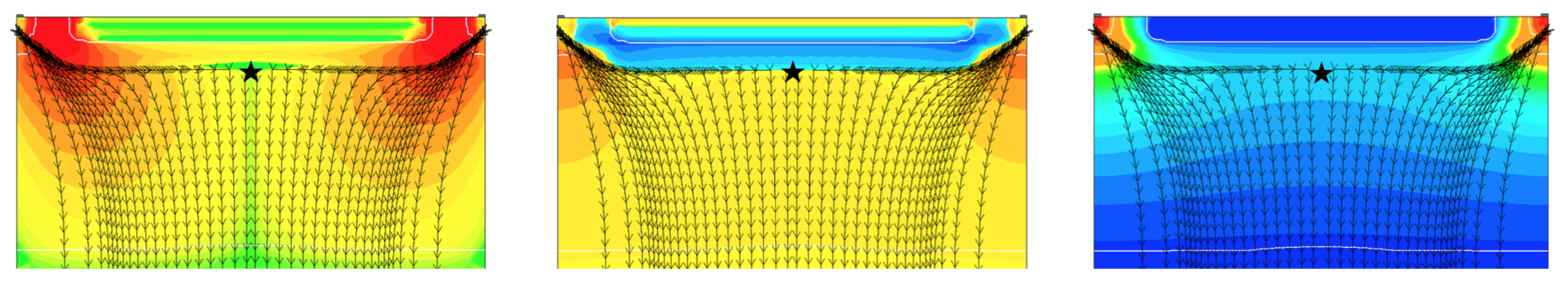}
	\put(2,80){\rotatebox{0}{ \scriptsize \textbf{Lateral electric field:}}}
	\put(150,80){\rotatebox{0}{ \scriptsize \textbf{Electric field along sensor depth:}}}
	\put(297,80){\rotatebox{0}{ \scriptsize \textbf{Electrostatic potential:}}}
\end{overpic}
\end{minipage}
\caption{Results of the electrostatic simulation for the modified process with a pixel size of $\mathrm{36.4 \mu m \times 36.4 \mu m}$.
The black arrows mark the electric field stream lines, the star symbol indicates the electric field minimum and the white lines mark the edges of the depleted regions.}
\label{fig:pot_mod}
\end{figure}

As shown in Figure~\ref{fig:pot_mod} for a pixel size of $\mathrm{36.4 \, \times \, 36.4 \, \mu m^2}$, the lateral electric field is due to symmetry zero 
at the pixel corners and the electric field along the sensor depth reaches a zero value close to the depth of the deep planar junction, resulting in a zero overall electric field at the pixel corners, a constant electrostatic potential, indicated by a star symbol in the figure.
As visualised by the black arrows, the direction of the electric field along the sensor depth results in a push of charge carriers created at various sensor depth at the pixel corner into this electric field minimum.
For the propagation of the charge out of this minimum the lateral component of the electric field is crucial.

\begin{figure}[!ht]
\centering
\begin{minipage}[t]{1\textwidth} 
\begin{overpic}[width=1\textwidth]{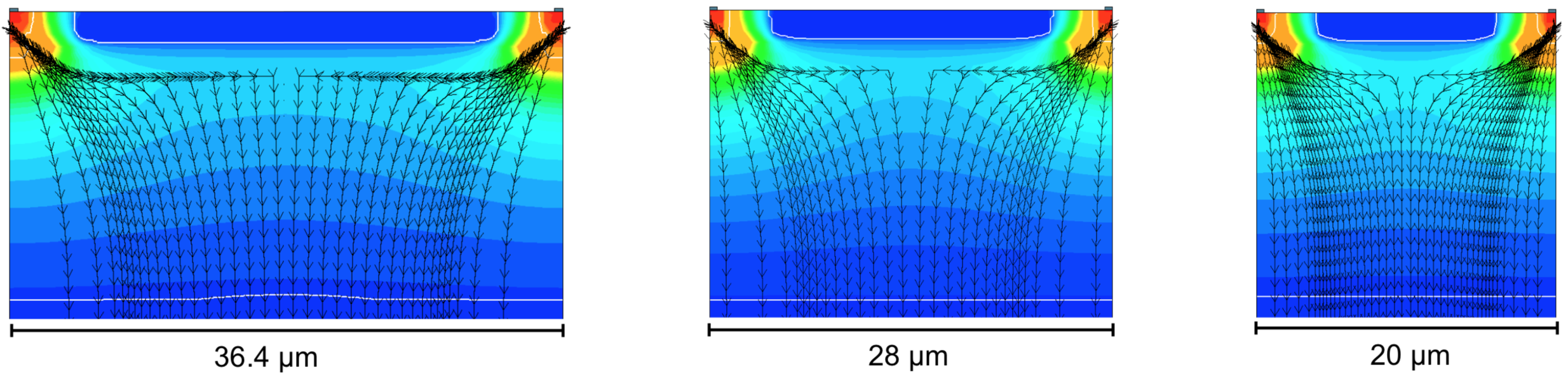}
\end{overpic}
\end{minipage}
\caption{Electrostatic potential for the modified process for different pixel sizes. 
The black arrows mark the electric field stream lines and the white lines mark the edge of the depleted regions.}
\label{fig:pot_mod_small_pixels}
\end{figure}

As shown in Figure~\ref{fig:pot_mod_small_pixels}, the size of the lateral field around the electric field minimum depends strongly on the pixel size:
The smaller the pixels, the larger the electrostatic potential difference and thus the electric field along the lateral dimension.
This helps to push the charge carriers out of this minimum and towards the collection electrode, as visualised by the electric field stream lines in Figure~\ref{fig:pot_mod_small_pixels}. 

\begin{figure}[!ht]
\centering
\vspace{0.1cm}
\begin{minipage}[t]{1\textwidth} 
\begin{overpic}[width=1\textwidth]{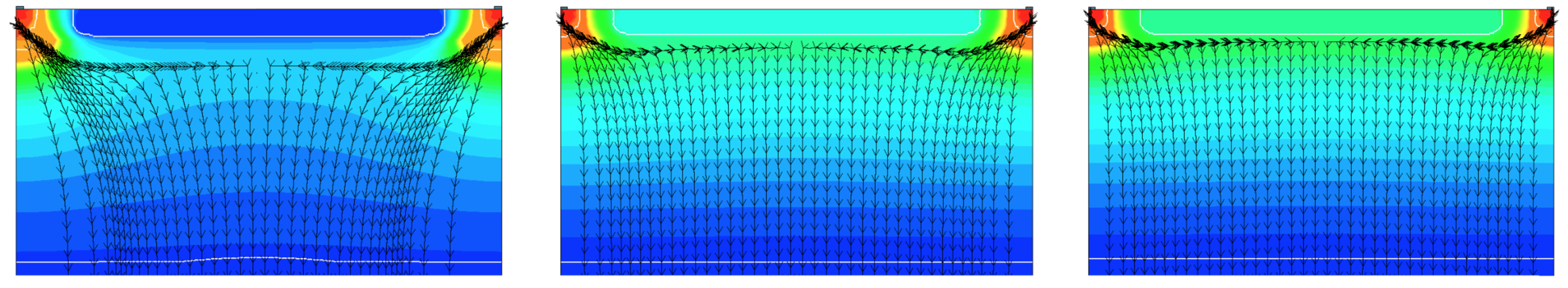}
\put(2,82){\rotatebox{0}{ \scriptsize \textbf{Backside voltage - 6 V:}}}
\put(151,82){\rotatebox{0}{ \scriptsize \textbf{Backside voltage - 15 V:}}}
\put(296,82){\rotatebox{0}{ \scriptsize \textbf{Backside voltage - 20 V:}}}
\end{overpic}
\end{minipage}
\caption{Electrostatic potential for different backside voltages for the modified process with a pixel size of $\mathrm{36.4 \mu m \times 36.4 \mu m}$.
The black arrows mark the electric field stream lines and the white lines mark the edges of the depleted regions.}
\label{fig:pot_dif_v}
\end{figure}

The importance of considering the direction of the electric field can be understood by inspecting different backside bias voltages for the modified process (see Figure~\ref{fig:pot_dif_v}).
For lower backside voltages the electric field along the sensor depth is decreased.
At the pixel corner, this results in a change of the direction of the electric field towards the collection electrode and thus a shorter drift path.

The electric field minimum results in a slower charge collection, creating a higher probability of charge trapping after irradiation.
The resulting dependency of the efficiency after irradiation on the pixel size has been observed in test-beam measurements~\cite{bojan,ivan,roberto,heinz_paper}.
Moreover, results with different p-well layouts have shown a higher efficiency in pixel regions where the p-well layout leads to a higher lateral field~\cite{bojan,roberto}.

Overall, the experimental results as well as the TCAD simulations indicate that increasing the lateral field is the key to increasing the charge collection to make CMOS sensors with a small collection electrode radiation hard and achieving precise timing resolution.  
While the pixel size is limited by the requirement to fit all needed circuitry, a change of the sensor concept is pursued in the following simulation studies to enhance the lateral field while only minimally changing the manufacturing process.

\section{Sensor concepts for a faster charge collection - three-dimensional electrostatic simulations}
\begin{figure}[!ht]
\centering
\vspace{0.4cm}
\begin{minipage}[t]{0.48\textwidth} 
\begin{overpic}[width=1\textwidth]{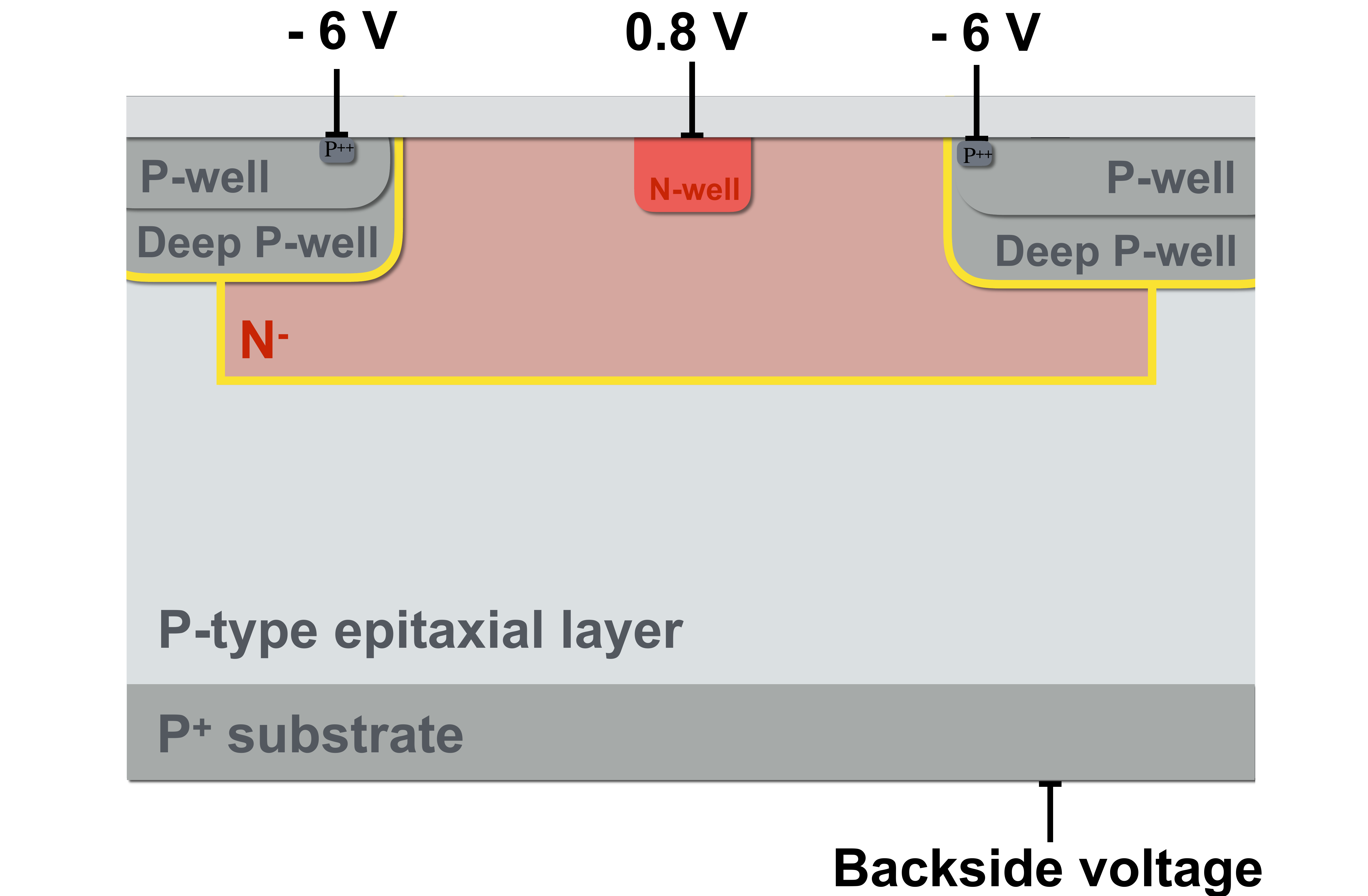}
	\put(18,139){\rotatebox{0}{ \scriptsize \textbf{Gap in deep n-implant:}}}
\end{overpic}
\end{minipage}
\hspace{0.3cm}
\begin{minipage}[t]{0.48\textwidth} 
\begin{overpic}[width=1\textwidth]{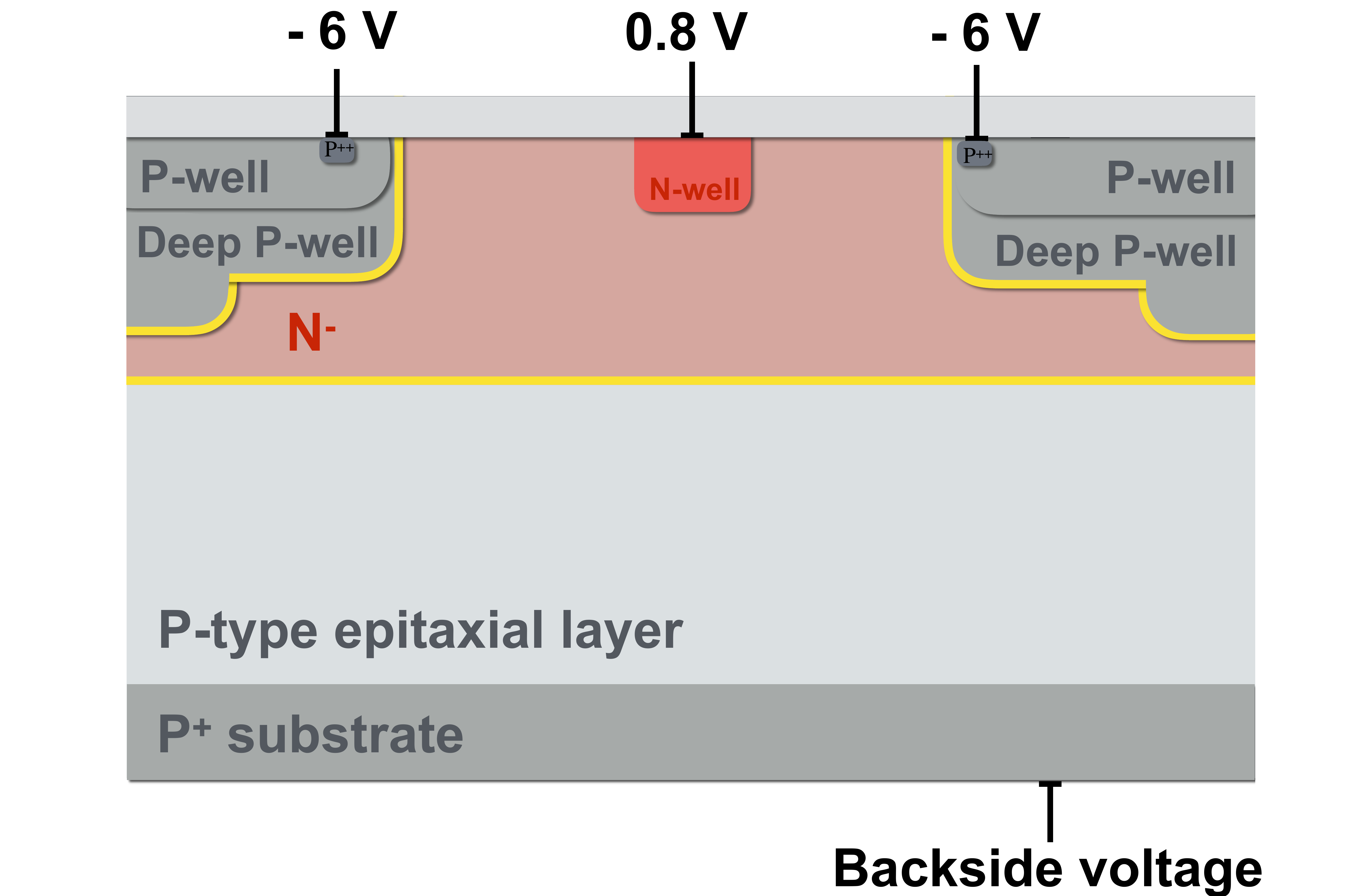}
		\put(18,139){\rotatebox{0}{ \scriptsize \textbf{Additional p-implant:}}}
\end{overpic}
\end{minipage}
\caption{Proposed concepts (not to scale) to increase the lateral electric field at the pixel borders: an additional p-implant (left) and gap in the deep n-implant (right). The yellow lines indicate the junctions.}
\label{fig:crosssection_mod}
\end{figure}

Figure~\ref{fig:crosssection_mod} shows two different approaches to increase the lateral electric field at the pixel borders: Creating a gap in the deep n-implant, requiring only a mask change, and introducing an additional p-type implant at the pixel border. 
Additional implants to accelerate the charge collection have also been pursued for image sensors for visible light detection~\cite{goji_paper} as well as for SOI sensors~\cite{soi_paper}.

\begin{figure}[!ht]
	\vspace{0.4cm}
\centering
\begin{minipage}[t]{1\textwidth} 
\begin{overpic}[width=1\textwidth]{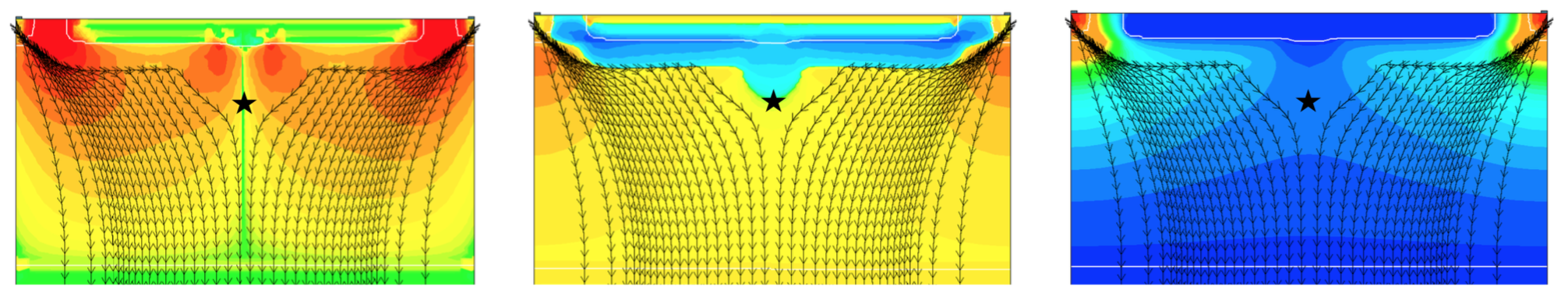}
	\put(2,87){\rotatebox{0}{ \scriptsize \textbf{Lateral electric field:}}}
	\put(144,87){\rotatebox{0}{ \scriptsize \textbf{Electric field along sensor depth:}}}
	\put(291,87){\rotatebox{0}{ \scriptsize \textbf{Electrostatic potential:}}}
\end{overpic}
\end{minipage}
\caption{Results of the electrostatic simulation for the concept with the gap in the deep n-implant with a pixel size of $\mathrm{36.4 \mu m \times 36.4 \mu m}$.
The black arrows mark the electric field stream lines, the star symbol indicates the electric field minimum and the white lines mark the edges of the depleted regions.}
\label{fig:pot_gap}
\end{figure}

\begin{figure}[!ht]
	\vspace{0.4cm}
\centering
\begin{minipage}[t]{0.99\textwidth} 
\begin{overpic}[width=1\textwidth]{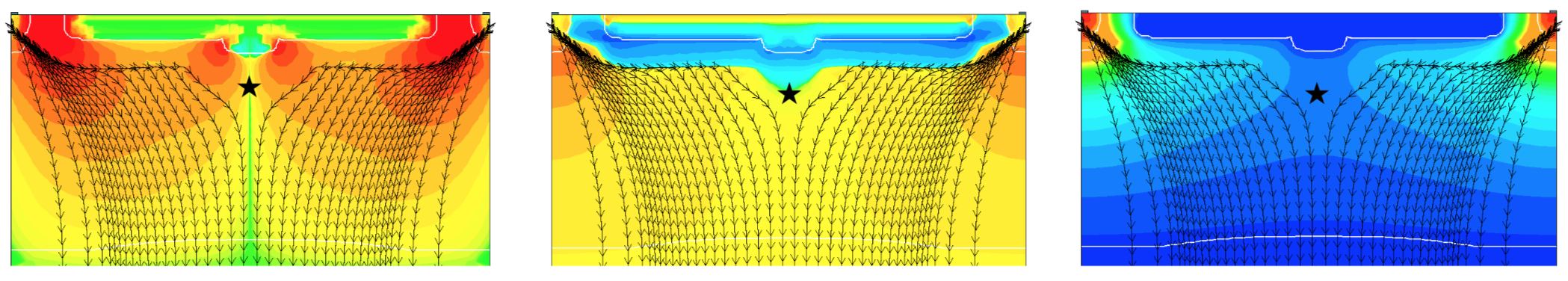}
	\put(2,80){\rotatebox{0}{ \scriptsize \textbf{Lateral electric field:}}}
	\put(147,80){\rotatebox{0}{ \scriptsize \textbf{Electric field along sensor depth:}}}
	\put(291,80){\rotatebox{0}{ \scriptsize \textbf{Electrostatic potential:}}}
\end{overpic}
\end{minipage}
\caption{Results of the electrostatic simulation for the concept with the additional p-implant with a pixel size of $\mathrm{36.4 \mu m \times 36.4 \mu m}$.
The black arrows mark the electric field stream lines, the star symbol indicates the electric field minimum and the white lines mark the edges of the depleted regions.}
\label{fig:pot_implant}
\end{figure}

Both approaches proposed here introduce a junction along the sensor depth, significantly increasing the lateral electric field, but also shifting the minimum of the electric field deeper into the silicon compared to the original approach shown in the right side of Figure~\ref{fig:process_mod}. As a result the electric field starts to bend towards the collection electrodes already deeper in the silicon, reducing the drift path and hence the charge collection time.
This is illustrated in Figure~\ref{fig:pot_gap} and Figure~\ref{fig:pot_implant} for the gap in the low-dose n-implant and the additional p-type implant, respectively. Cuts through the pixel centre of the simulated three dimensional pixel cell are presented for a simulation with $\mathrm{0.8\, V}$ collection electrode bias and $\mathrm{-\, 6\, V}$ bias on p-wells and substrate with a pixel size of $\mathrm{36.4 \,\mu m \time 36.4 \, \mu m}$.

\section{Transient three-dimensional TCAD simulations }

In the previous section the influence of the pixel size and two additional pixel modifications on the electric field was illustrated using electrostatic simulations.
To compare the timing response for different cases, three-dimensional transient TCAD simulation results are presented for a Minimum Ionising Particle (MIP) traversing the pixel corner, the worst case in terms of charge collection time. Results are shown both, non-irradiated sensors and for sensors irradiated with a fluence of $\mathrm{10^{15} neq /cm^2}$. To model the effect of radiation damage, defect levels have been introduced, as described in~\cite{paper_irrad}.
In the following the influence of the pixel modifications, of the pixel size, and of the sensor reverse backside bias are discussed.
The voltage on the collection electrode and p-wells has been set to $\mathrm{0.8 \, V}$ and $\mathrm{- \, 6 \, V}$, respectively.

\subsection{Pixel modifications}

\begin{figure}[!ht]
\centering
\vspace{0.2cm}

\begin{minipage}[t]{0.4\textwidth} 
\begin{overpic}[width=1\textwidth]{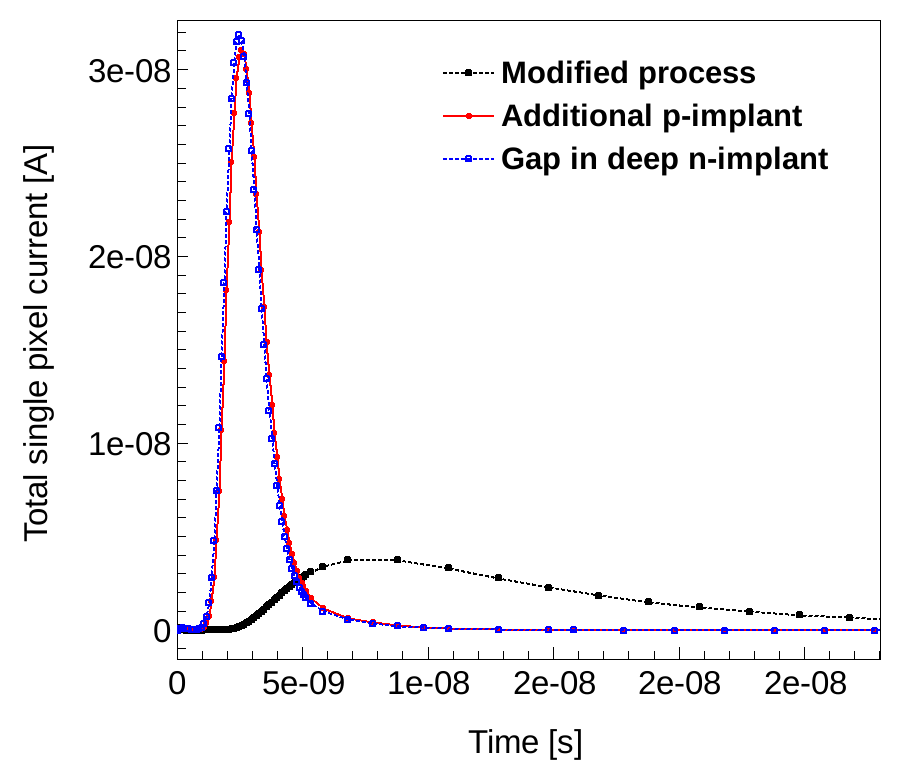}
	\put(33,152){\rotatebox{0}{ \scriptsize \textbf{Before irradiation:}}}
\end{overpic}
\end{minipage}
\hspace{0.4cm}
\begin{minipage}[t]{0.4\textwidth} 
\begin{overpic}[width=1\textwidth]{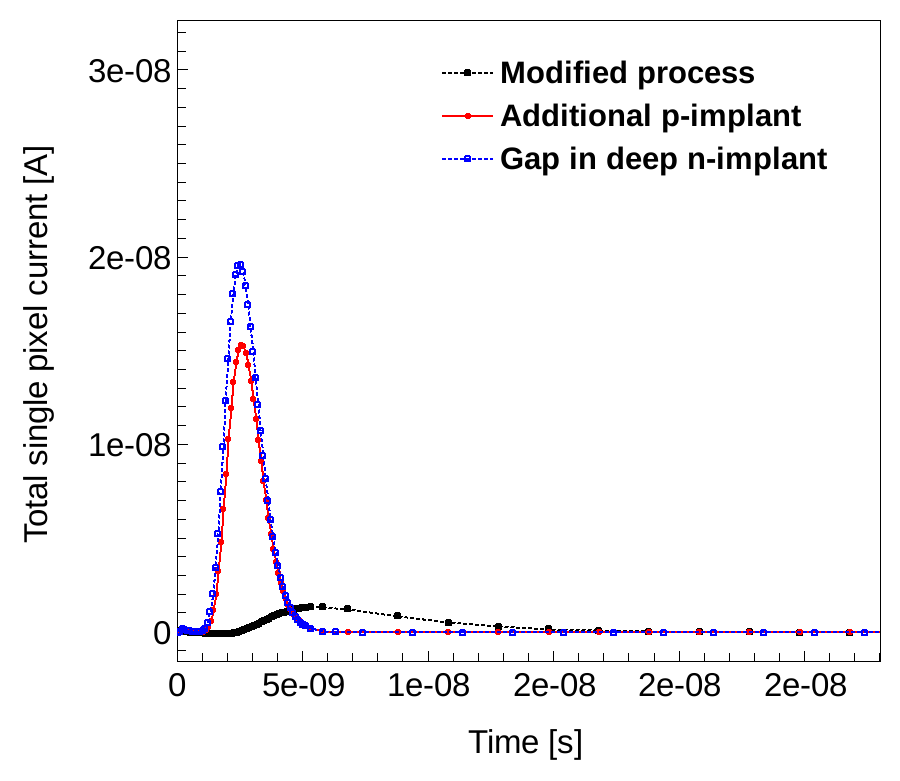}
	\put(33,152){\rotatebox{0}{ \scriptsize \textbf{After irradiation:}}}
	\put(112,100){\rotatebox{0}{ \scriptsize Fluence of}}
	\put(112,88){\rotatebox{0}{ \scriptsize  $\mathrm{10^{15} neq /cm^2}$}}
\end{overpic}
\end{minipage}
\hspace{0.4cm}	
\caption{Current versus time for different sensor concepts with a pixel size of $\mathrm{36.4 \mu m \times 36.4 \mu m}$, simulating a MIP incident at the pixel corner. A significantly faster charge collection has been simulated for the additional p-implant and the gap in the deep n-implant (coloured lines) compared to the modified process (black).}
\label{fig:current_pulse}
\end{figure}

The current induced on a single pixel is presented versus time in Figure~\ref{fig:current_pulse} for the different sensor concepts before (left) and after (right) irradiation for a backside voltage of $\mathrm{- \, 6 \, V}$.
The charge collection time is reduced by a factor of at least two for the proposed concepts.
The same general trends can be observed after irradiation.
However, the overall pulse heights are significantly reduced, as explained by trapping and recombination of the charge carriers.

\begin{figure}[!ht]
\centering
\vspace{0.2cm}
\begin{minipage}[t]{0.4\textwidth} 
\begin{overpic}[width=1\textwidth]{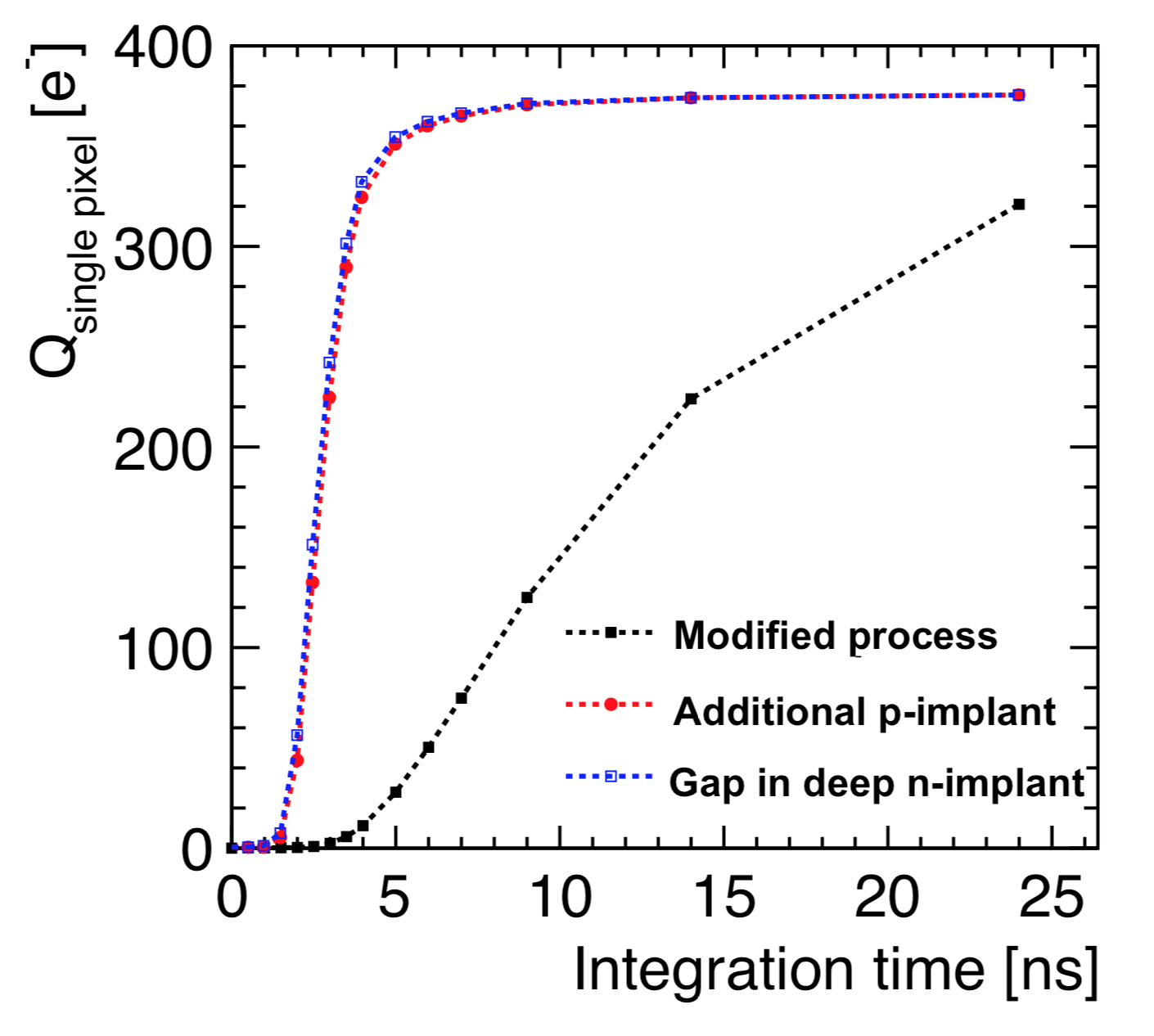}
	\put(33,152){\rotatebox{0}{ \scriptsize \textbf{Before irradiation:}}}
\end{overpic}
\end{minipage}
\hspace{0.4cm}	
\begin{minipage}[t]{0.4\textwidth} 
\begin{overpic}[width=1\textwidth]{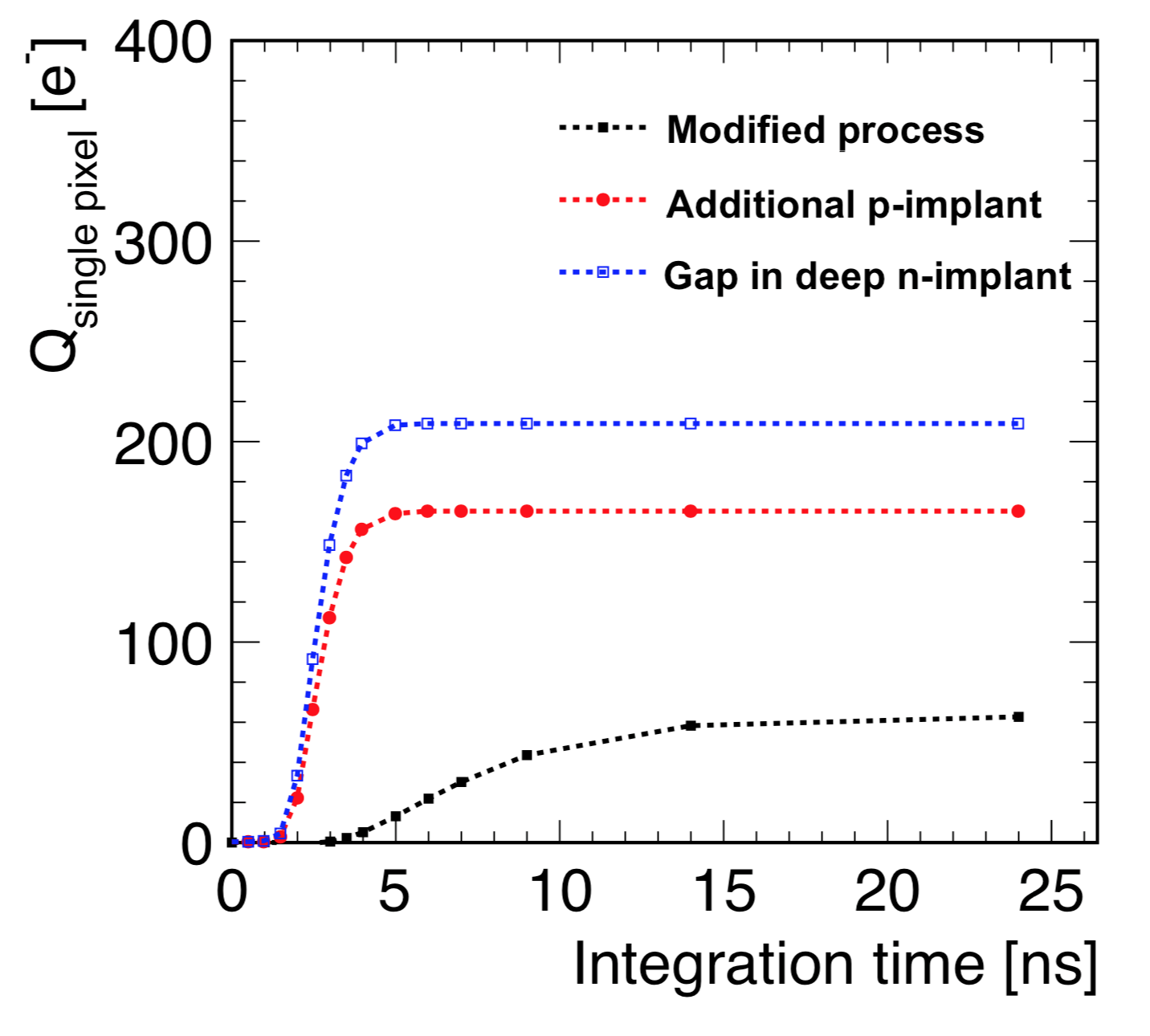}
	\put(33,152){\rotatebox{0}{ \scriptsize \textbf{After irradiation:}}}
	\put(112,65){\rotatebox{0}{ \scriptsize Fluence of}}
	\put(112,53){\rotatebox{0}{ \scriptsize $\mathrm{10^{15} neq /cm^2}$}}
\end{overpic}
\end{minipage}
\caption{Collected charge versus integration time for different sensor concepts with a pixel size of $\mathrm{36.4 \mu m \times 36.4 \mu m}$ before (left) and after (right) irradiation. }
\label{fig_charge}
\end{figure}

The differences in pulse height have been evaluated by integrating the current pulses and calculating the charge.
Figure~\ref{fig_charge} shows the charge versus integration time before (left) and after (right) irradiation.
Differences are already observable before irradiation:
While most of the charge is collected for the concept with the additional p-implant and the gap in the deep n-implant, not all charge is collected for the modified process within $\mathrm{25 \, ns}$.
This illustrates the need of a process modification for faster charge collection even without irradiation for applications with a short integration time. Both proposed pixel improvements increase the collected charge after irradiation by at least a factor of three.

\subsection{Pixel size}

Moving towards smaller feature sizes will allow smaller pixel sizes while maintaining functionality. 
Thus, to evaluate the future prospects of the proposed sensor design concepts, the modified process and the concept with the additional p-implant are compared for smaller pixel sizes after irradiation.

\begin{figure}[!ht]
\centering
\vspace{0.6cm}
\begin{minipage}[t]{0.45\textwidth} 
\begin{overpic}[width=1\textwidth]{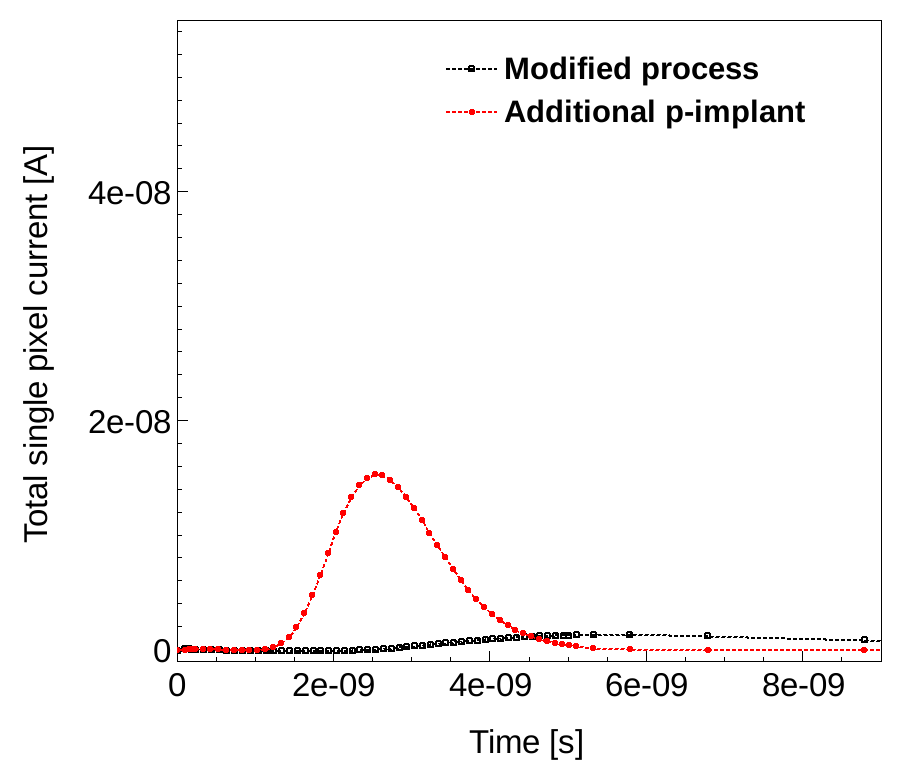}
	\put(35,168){\rotatebox{0}{ \scriptsize \textbf{Pixel size $\mathrm{\bold{36.4 \mu m \times 36.4 \, \mu m}}$:}}}
\end{overpic}
\end{minipage}
\hspace{0.1cm}
\begin{minipage}[t]{0.45\textwidth} 
\begin{overpic}[width=1\textwidth]{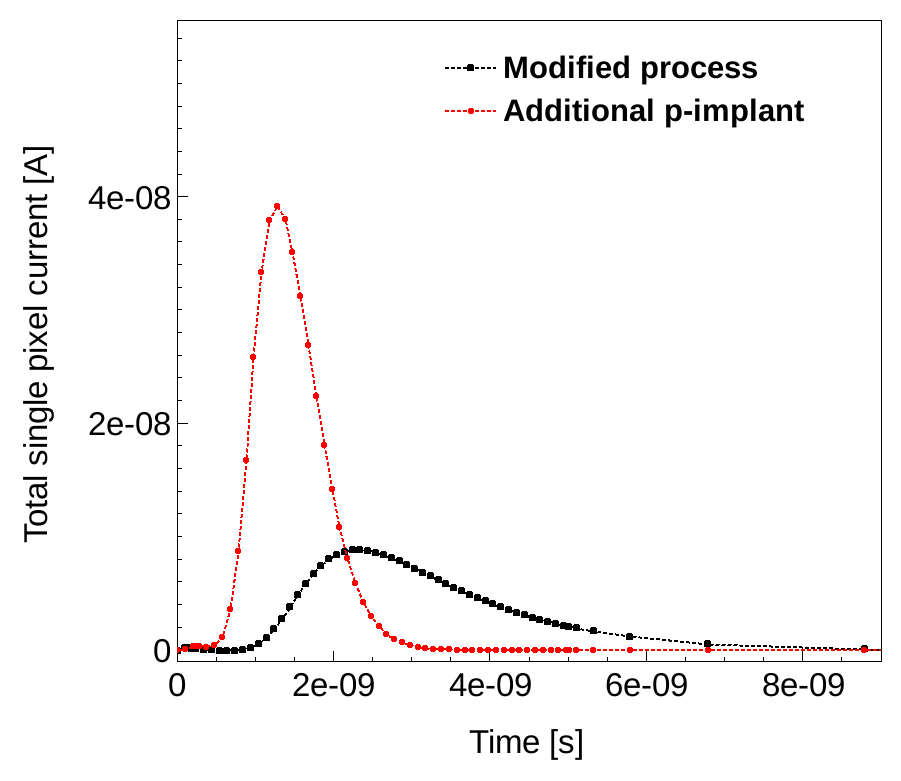}
	\put(35,168){\rotatebox{0}{ \scriptsize \textbf{Pixel size $\mathrm{\bold{28 \mu m \times 28 \, \mu m}}$:}}}
\end{overpic}
\end{minipage}\\

\vspace{0.5cm}
\begin{minipage}[t]{0.45\textwidth} 
\begin{overpic}[width=1\textwidth]{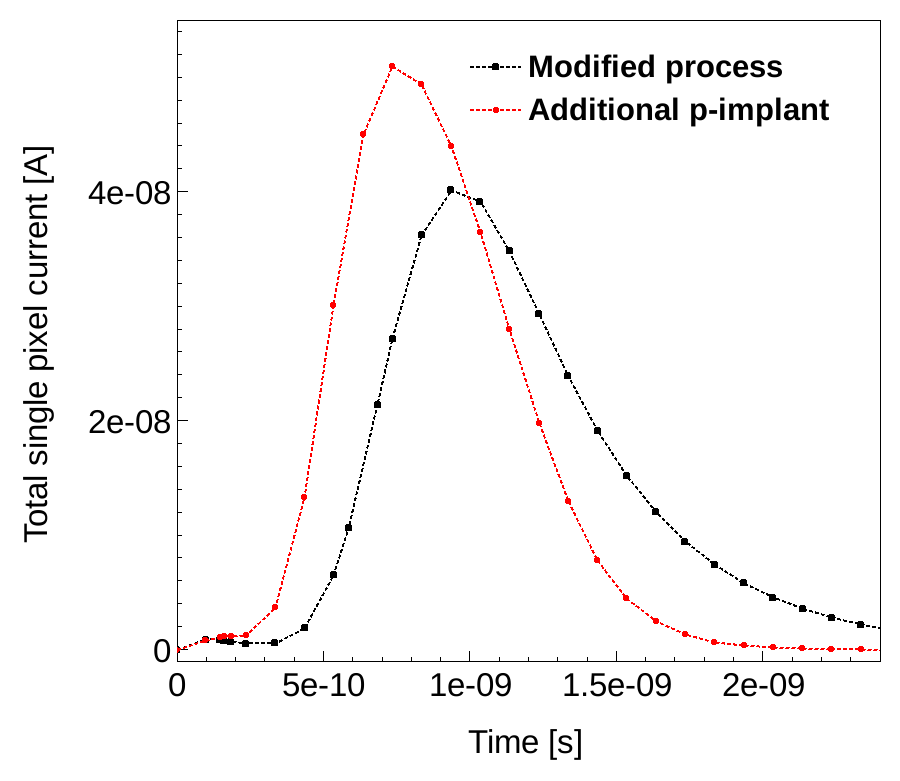}
	\put(35,168){\rotatebox{0}{ \scriptsize \textbf{Pixel size $\mathrm{\bold{20 \mu m \times 20 \, \mu m}}$:}}}
\end{overpic}
\end{minipage}
\hspace{0.1cm}
\caption{Current pulses simulating a MIP incident at the pixel corner for the modified process (black) and the concept with the additional p-implant compared for different pixel sizes after irradiation with a fluence of $\mathrm{10^{15} neq /cm^2}$. Note the different scale of the x-axis for a pixel size of $\mathrm{20 \, \mu m \, \times \, 20 \, \mu m}$.}
\label{fig:pulse_small_pixel}
\end{figure}

\begin{figure}[!ht]
\centering
\vspace{0.4cm}

\begin{minipage}[t]{0.47\textwidth} 
\begin{overpic}[width=1\textwidth]{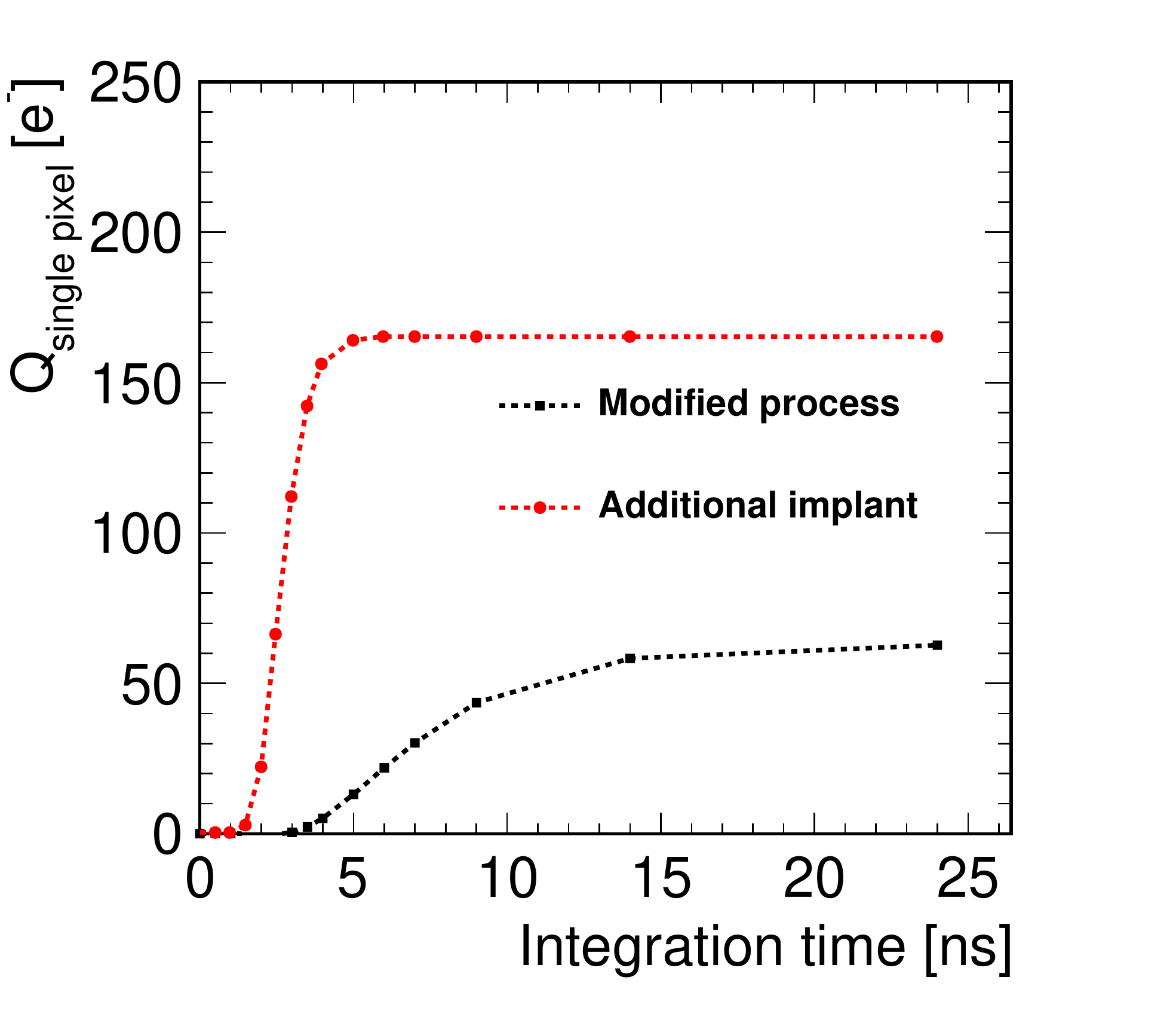}
	\put(32,165.5){\rotatebox{0}{ \scriptsize \textbf{Pixel size $\mathrm{\bold{36.4 \mu m \times 36.4 \, \mu m}}$:}}}
\end{overpic}
\end{minipage}
\begin{minipage}[t]{0.47\textwidth} 
\begin{overpic}[width=1\textwidth]{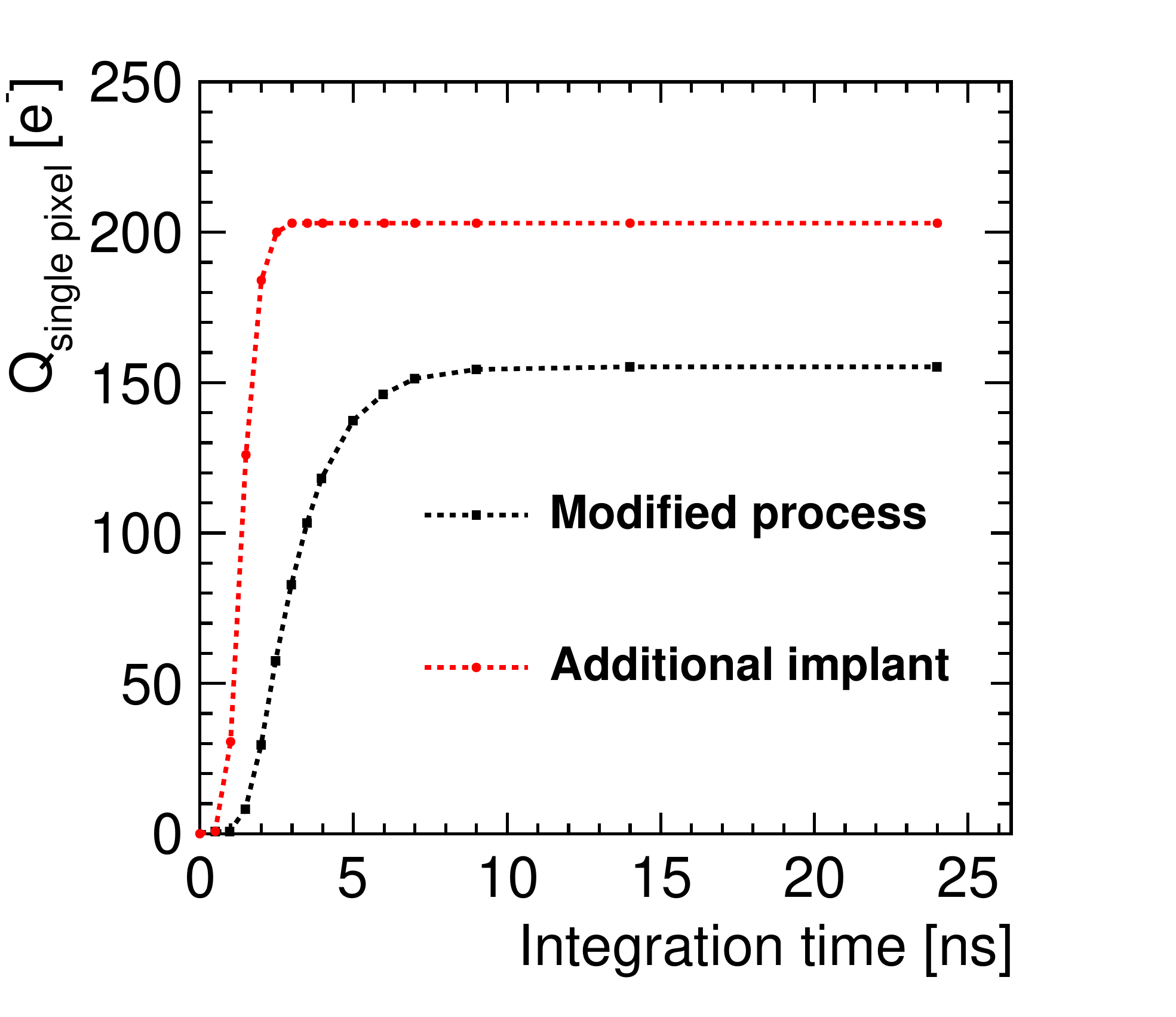}
	\put(32,165.5){\rotatebox{0}{ \scriptsize \textbf{Pixel size $\mathrm{\bold{28 \mu m \times 28 \, \mu m}}$:}}}
\end{overpic}
\end{minipage}\\
\vspace{0.4cm}
\begin{minipage}[t]{0.47\textwidth} 
\begin{overpic}[width=1\textwidth]{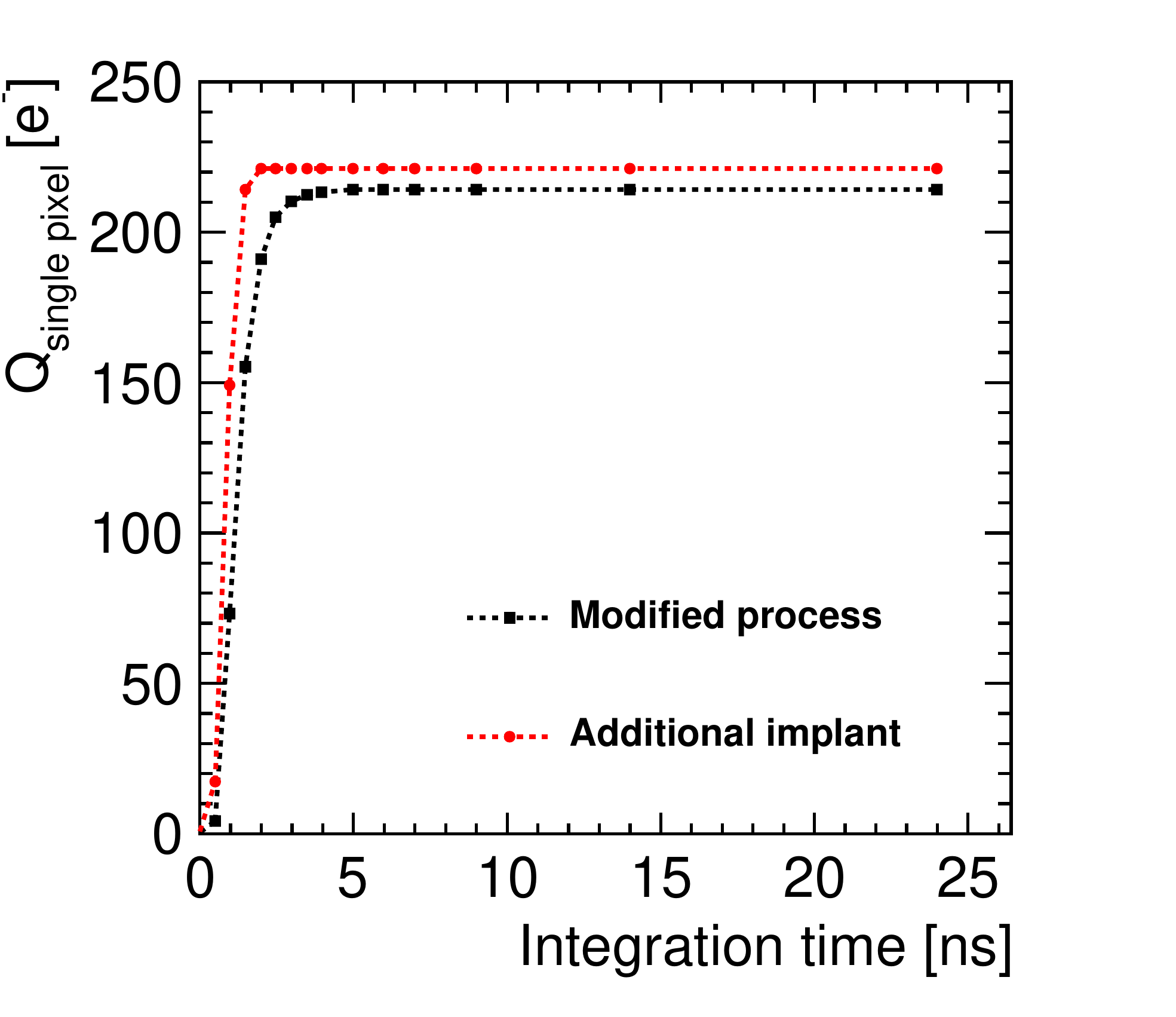}
	\put(32,165.5){\rotatebox{0}{ \scriptsize \textbf{Pixel size $\mathrm{\bold{20 \mu m \times 20 \, \mu m}}$:}}}
\end{overpic}
\end{minipage}
\caption{Collected charge versus integration time for the modified process (black) and the concept with the additional p-implant (red) compared for different pixel sizes after irradiation with a fluence of $\mathrm{10^{15} neq /cm^2}$.}
\label{fig:charge_t}
\end{figure}

Current pulses are presented for a backside voltage of $\mathrm{- \, 6 \, V}$ for different pixel sizes in Figure~\ref{fig:pulse_small_pixel}, comparing the original modified process from Figure~\ref{fig:process_mod}, with the concept with the additional p-implant in Figure~\ref{fig:crosssection_mod}.
Even for small pixel sizes of $\mathrm{20 \, \mu m \, \times \, 20 \, \mu m}$ the additional p-type implant significantly accelerates the charge collection, resulting in sub-nanosecond peaking times.
The charge versus integration time for different pixel sizes presented in Figure~\ref{fig:charge_t}, shows that the charge lost after irradiation can be recovered by going to smaller pixel sizes as well as by the proposed sensor modifications.

\subsection{Sensor reverse bias}

The maximal reverse bias voltage applicable to the p-wells is limited by the CMOS circuitry to $\mathrm{- \, 6 \, V}$~\cite{jacobus_thesis}.
The deep low-dose n-implant isolates the p-wells from the backside substrate and allows for a higher reverse bias on the substrate.
The two pixel improvements weaken this isolation, resulting in a high current flow between the p-wells and the backside substrate (punch-through). 
This is further investigated here by fixing the collection electrode and p-well bias to $\mathrm{0.8}$ and $\mathrm{-6 \, V}$ respectively, and sweeping the substrate bias from $\mathrm{0 \, V}$ to $\mathrm{- 20 \, V}$.

\begin{figure}[!ht]
\centering
\begin{minipage}[t]{0.48\textwidth} 
\begin{overpic}[width=1\textwidth]{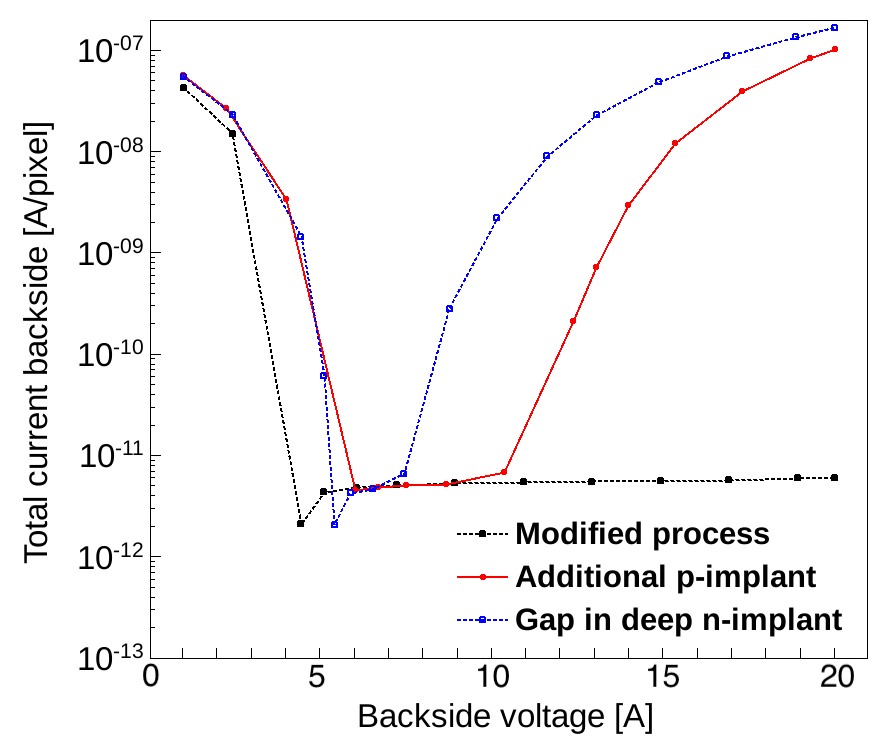}
\end{overpic}
\caption{Current flow between the p-wells and the backside for the different sensor concepts with a size of $\mathrm{36. 4 \mu m \times 36.4 \mu m}$.}
\label{fig:punch}
\end{minipage}
\hspace{0.3cm}
\begin{minipage}[t]{0.48\textwidth} 
\begin{overpic}[width=1\textwidth]{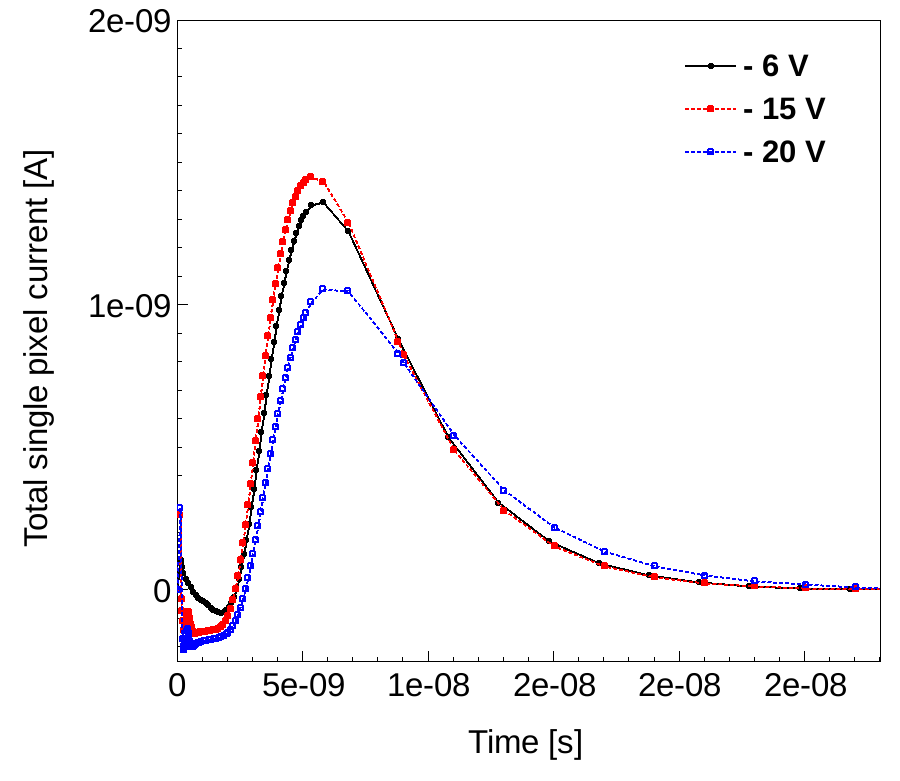}
\end{overpic}
\caption{Current pulse simulating a MIP incident at the pixel corner for different backside voltages for the modified process with a pixel size of $\mathrm{36. 4 \mu m \times 36.4 \mu m}$ after irradiation with a fluence of $\mathrm{10^{15} neq /cm^2}$.}
\label{fig:pulse_diff_backbias}
\end{minipage}
\end{figure}

For each step of the backside voltage the current flow between the backside and the p-wells has been calculated, as presented in Figure~\ref{fig:punch}.
A high current flow is observable for backside voltages below the $\mathrm{- \, 6 \, V}$ applied to the p-wells, since the small depletion of the epitaxial layer does not sufficiently isolate the p-wells from the backside, resulting in punch through between the p-wells and the backside substrate.
For backside voltages higher than the $\mathrm{- \, 6 \, V}$ applied to the p-wells, the modified process shows the expected isolation.
For the additional p-implant and the gap in the deep n-implant this isolation is reduced to a smaller voltage range and minimal for the sensor concept with the gap in the deep n-implant, leading to punch-through at lower absolute bias voltages.

A simulation of a MIP traversing the pixel corners has been performed for the modified process applying higher backside voltages, to investigate the impact on the charge collection time.
The current pulses after irradiation are presented in Figure~\ref{fig:pulse_diff_backbias}.

In the pixel corners a slight improvement can be noted for a backside voltage of $\mathrm{- \, 15 \, V}$.
An even higher backside voltage of $\mathrm{- \, 20 \, V}$ reduces the pulse height and thus the amount of collected charge, as explained by the higher electric field along the sensor depth that results in a longer drift path, a slower charge collection and a higher recombination probability after irradiation (see Figure~\ref{fig:pot_dif_v}).

\section{Summary}
By combining the advantages of a small sensor capacitance and a fully monolithic technology, CMOS pixel sensors with a small collection electrode address the requirements of future experiments.
However, experimental evidence showed that after irradiation signal charge was lost at the pixel corners causing severe detection inefficiencies even after a process modification to fully deplete the epitaxial layer.
Three-dimensional electrostatic TCAD simulations identified an electric field minimum at the pixel corners increasing the charge collection time and thus the probability of charges to be trapped after irradiation. 
Two different further sensor modifications were presented to reduce this electric field minimum and accelerate the collection of signal charge from the pixel edge towards the collection electrode.
This not only reduces the probability of the signal charge to be trapped but simultaneously improves the precision of the time stamping capability.
Three dimensional transient TCAD simulations show these sensor modifications indeed accelerate the charge collection time by approximately a factor four. This gives confidence that the post irradiation performance will be improved, and the post-irradiation simulations confirm this: the amount of collected charge after irradiation with a fluence of $\mathrm{10^{15} neq /cm^2}$ has been increased by a factor of approximately three in simulations for a $\mathrm{36.4 \mu m}$ pixel pitch, with an additional further improvement for smaller pixel pitches. The post-irradiation models taken from literature have not been specifically developed and tuned for epitaxial material limiting the quantitative precision of post-irradiation predictions. However, the underlying concept of accelerating the charge collection and thereby decreasing the recombination probability is not dependent on the irradiation model.


\end{document}